\documentclass[12pt,conference]{IEEEtran}
\usepackage{graphicx}

\begin{document}
\title{Exhaustive Verification of Weak Reconstruction For Self Complementary Graphs }
\author{{ Prof. S. K. Gupta} \and { Sahil Singla} \and { Akash Khandelwal}\and { Srilekha} \and { Apurv Tiwari }}
%\author{\IEEEauthorblockN{S. K. Gupta, Sahil Singla, Akash Khandelwal, Apurv Tiwari} \IEEEauthorblockA{IIT Delhi\\ \{skg, cs1070181, cs5070208, apurv\}@cse.iitd.ac.in} \and \IEEEauthorblockN{Srilekha} \IEEEauthorblockA{NIT Karnatka\\ srilekha.vk@gmail.com} }
\author{\IEEEauthorblockN{S. K. Gupta , Sahil Singla, Akash Khandelwal, Apurv Tiwari, Srilekha} \IEEEauthorblockA{ \{skg, cs1070181\}@cse.iitd.ac.in, \{akash3119, apurvtwr, srilekha.vk\}@gmail.com}  }

\maketitle

\begin{abstract}
This paper presents an exhaustive approach for verification of the weak reconstruction of Self Complementary Graphs upto $17$ vertices. It describes the general problem of the Reconstruction Conjecture, explaining the complexity involved in checking deck-isomorphism between two graphs. In order to improve the computation time, various pruning  techniques have been employed to reduce the number of graph-isomorphism comparisons. These techniques offer great help in proceeding with a reconstructive approach. An analysis of the numbers involved is provided, along with the various limitations of this approach. A list enumerating the number of SC graphs up till $101$ vertices is also appended.  
\end{abstract}

\begin{figure*}
	  \centering 
           \includegraphics[width=18cm]{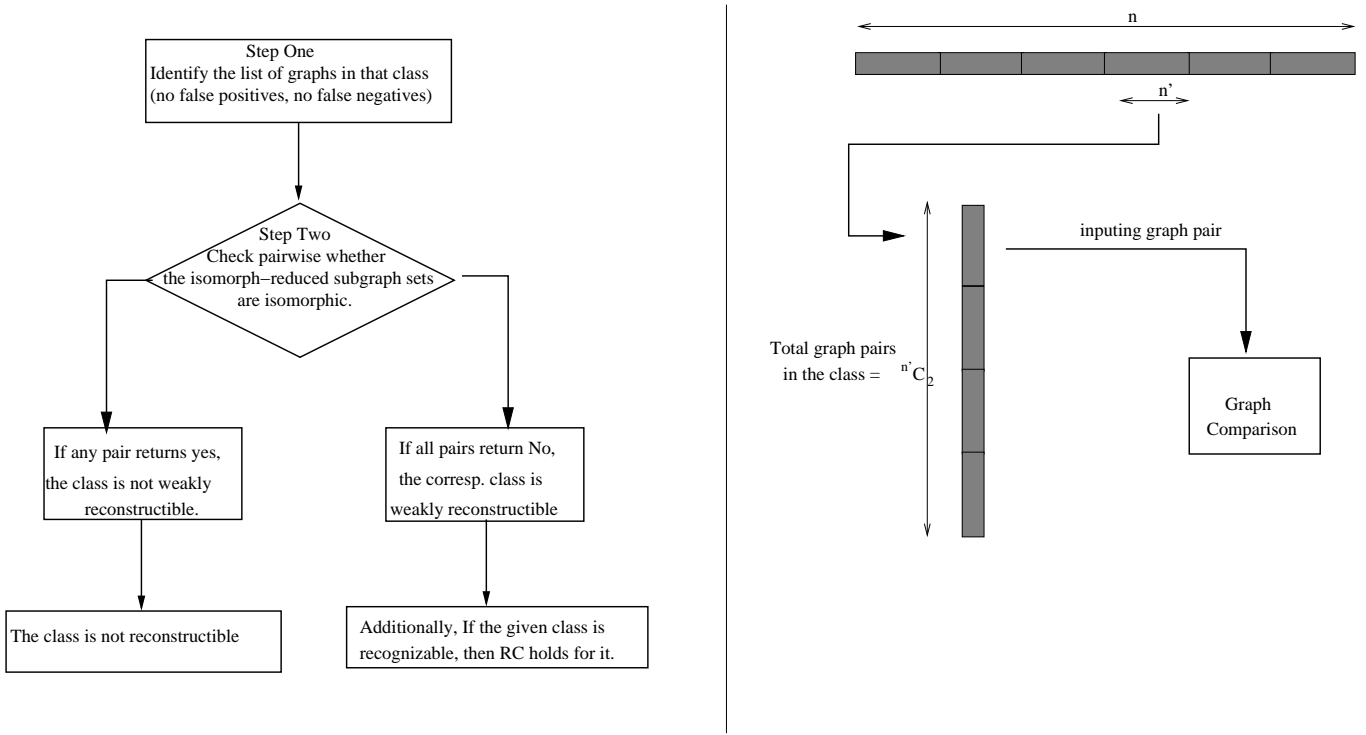}
	   \caption{The Procedure for RC Verification, depicted as (i)Overall Flowchart (ii)Line Diagram for the Approach Employed}
	   \label{flowchart}
  \end{figure*}

\section{Introduction}
\subsection{ The Reconstruction Conjecture (RC)}
\paragraph*{} The Reconstruction Conjecture (RC) is one of the most celebrated unsolved problems in Discrete Mathematics and Combinatorics circles. It was first discovered by S.M.  Ulam and P. J. Kelly in 1941 \cite{b3} . Any graph $G$ has a vertex set $V(G)$ and an edge set $E(G)$. A vertex-deleted-subgraph of $G$, $G_i$, is the unlabelled subgraph of $G$ with the $i^{th}$ vertex and its coincident edges removed. The deck of the graph $G$ is the collection of all vertex-deleted subgraphs of $G$. For terms not defined here, we shall use the terminology followed in Harary \cite{h1}. 

\subsubsection{ Original Definition}
\paragraph*{} Ulam \cite{u1} states the following problem:
\\
``Suppose that in two sets $A$, $B$; each of $n$ elements, there is defined a distance function $\rho$ for every pair of distinct points, with values either $1$ or $2$ and $\rho(x,x) =0$. Assume that for every subset of $n-1$ points of $A$; there exists an  isometric system of $n-1$ points of $B$, and that the number of distinct subsets isometric to any given subset of $n-1$ points  is same in $A$ as in $B$. Are $A$ and $B$ isometric?''

\subsubsection{ Modified Definition of the Graph Reconstruction Conjecture}
Reconstruction Conjecture \cite{h1}: ``A simple finite graph $G$ with at least three points can be reconstructed uniquely (up to isomorphism) from its collection of vertex deleted subgraphs $G_i$.''  This conjecture was called by Harary  \cite{h1},  a ''graphical disease'', along with the 4-Color Conjecture and the characterization of Hamiltonian graphs.

\subsection{ Reconstructive Approach Towards RC }
\paragraph*{} The reconstruction problems provide a fascinating study of the structure of graphs. The identification of structure of a graph is the first step in its reconstruction. We can determine various invariants of a graph from its subgraphs, which in turn tell us about the structure of the graph.
\paragraph*{} One of the ways for tackling the RC is known as the reconstructive approach, and is followed in many of the proofs of the conjecture for specific classes. This approach relies on two parts: one- ``recognizability``, and two – ''weak reconstructibility``\cite{b3}. The class of graphs $C$ to which $G$ belongs is said to be recognizable if every reconstruction of $G$ belongs to the class $C$. The class $C$ is said to be weakly reconstructible if every reconstruction of a graph $G$ belonging to the class $C$ is isomorphic to $G$, for each $G$ in $C$ \cite{b3}.
\paragraph*{} A parameter of  $G$  which can be deduced from the deck is called reconstructible.  Another approach attempted by many is in determining reconstructibility of a graph on the basis of their graph invariants.  Various  properties such as  characteristic polynomial \cite{e1}, degree sequence \cite{w1} and diameter \cite{s2} have been proven to be reconstructible for SC graphs.  Additionally, graph invariants like number of vertices, edges, blocks, cut-vertices, independent cycles and connectivity  have been proven to be reconstructible \cite{k2}. 

\subsection{ Discussion about the Problem }
\paragraph*{} McKay\cite{b1} employed an exhaustive technique to show that graphs up to  $11$ vertices are determined uniquely by their sets of vertex-deleted subgraphs, even if the   set of subgraphs is reduced by isomorphism type.  
\paragraph*{} The statement of the conjecture excludes the trivial graph $K1$, graphs on two vertices and infinite graphs. The deck of graphs on two points, i.e. $K2$ and $K2'$,are clearly homomorphic (a pair of $K1$s comprising each of their decks) but the graphs are non-isomorphic. For every infinte cardinal $\alpha$, there exists a graph with $\alpha$ edges which is not uniquely reconstructible from its family of edge deleted sub-graphs \cite{c1} .
%A large number of classes of infinite graphs (tree on infinite vertices) that have been proved to be non-reconstructible. One of the appreciable facts is that each of the graphs in the non-reconstructible pair is isomorphic to an induced subgraph of the other[reference].
 Apart from these two exceptions which prohibit the conjecture from encompassing all graphs, unique reconstructibility is conjectured for all other graphs.
\paragraph*{} The conjecture has been proved for a number of infinitely sized classes of graphs, such as trees \cite{k1}, squares of trees \cite{s1}, unicyclic graphs \cite{m2}, regular graphs \cite{n2} and disconnected graphs \cite{h2}. Though the problem can be stated very simply,yet due to a lack of a nice set of characterizing invariants, it has still not been proven for very important classes of graphs like bipartite graphs \cite{b3} and planar graphs \cite{b3}. For further study of this problem, the reader is referred to survey by Bondy \cite{b3}. 
\paragraph*{} In a probabilistic sense, it has been shown that almost all graphs are reconstructible \cite{b2}. This means that the probability that a randomly chosen graph on  $n$  vertices is not reconstructible goes to $0$ as  $n$  goes to infinity. In fact, it was shown that not only are almost all graphs reconstructible, but also that the entire deck is not generally necessary to reconstruct them — almost all graphs have the property that there exist three cards in their deck that uniquely determine the graph.

\begin{figure*}
	  \centering 
           \includegraphics[width=18cm]{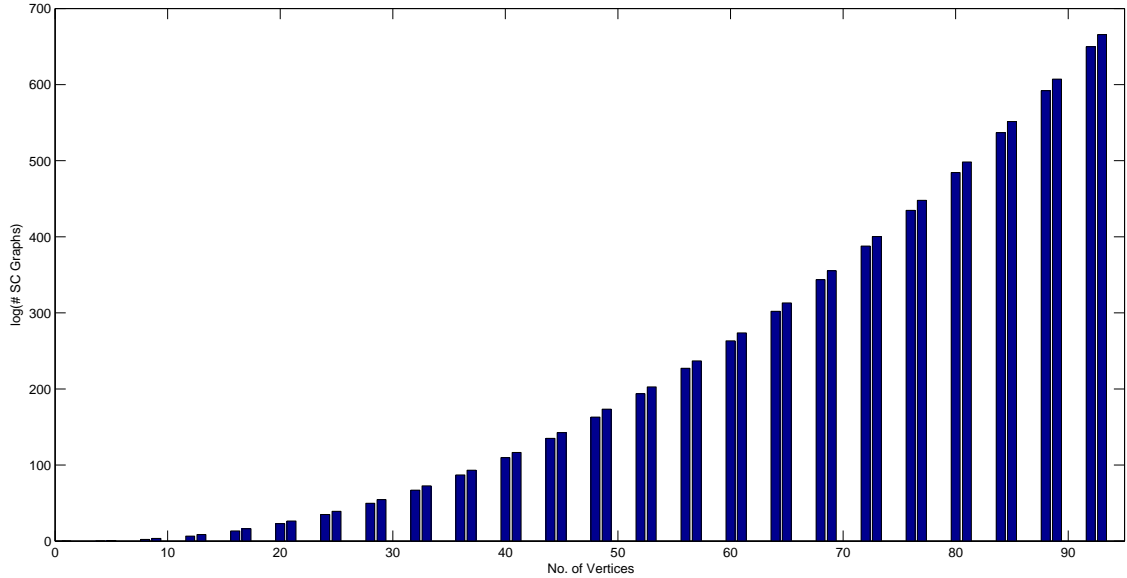}
	   \caption{A graph showing logarithm of No. of SC Graphs vs No. of vertices}
	   \label{plot}
  \end{figure*}

\section{ RC Verification for Self-Complementary Graphs}

\subsection{\label{2.1} Overview of Procedure Employed }
\paragraph*{} The problem definition refers to unlabeled graphs making it computationally expensive(refer Sec.\ref{2.2} for the worst case analysis). An idea of the numbers that are associated with the problem, is presented in Sec.\ref{2.4} The appendix lists the variation in the number of SC graphs with the number of vertices. 
\paragraph*{} This paper suggests techniques through use of efficient structures for storage and suitable classifications(described in Sec.\ref{2.5}). The intelligence of such techniques lies in reducing the complexity of the underlying problem.

\begin{figure*}
	  \centering 
           \includegraphics[width=18cm]{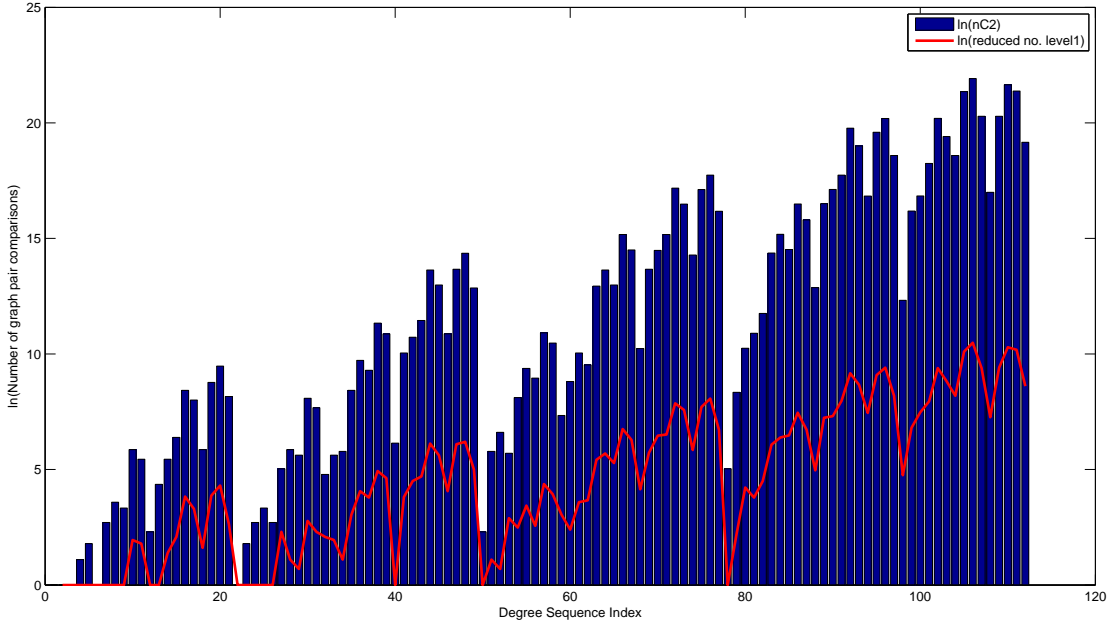}
	   \caption{A graph showing reduction in the number of comparisons after Level-1 Pruning}
	   \label{prune}
  \end{figure*} 

\paragraph*{} For detailed a survey of self complementary graphs, reader is referred to \cite{f2}. The listing  of self-complementary graphs is available upto 17 vertices \cite{b6}. This listing of self complementary graphs is generated using the fact that every self complementary graph ($G$) on $4n$ vertices can be broken down into edge-disjoint subgraphs $H$, $H^*$ and $B$, such that $G = H+B+ H^*$ , $H' = H^{*}$ and $B$ is a bipartite graph between the vertex sets of $H$ and $H^{*}$. There exists a self-complementing permuation (sigma) with even length  cycles. This permutation plays a key role in generating the self complementary graphs \cite{n3}. An efficient way of generating SC graphs on $4n+1$ vertices using the set of SC graphs on $4n$ vertices is discussed in \cite{x1}. Although this process is exhaustive, it creates multiple copies of a graph in the form of isomorphic graphs. In \cite{m3} , a method is suggested to reduce the generation of such copies of graphs. In this procedure, the permutations of vertices within a cycle of a self-complementing permutation are avoided as each such permutation generates the same set of graphs.
\paragraph*{} In proving that self complementary graphs up to $17$ vertices are uniquely determined (within the set  of  all graphs) by their decks, the algorithmic challenge lies in reducing the number of comparisons among graph-pairs. The flowchart of the procedure adopted is in Fig. \ref{flowchart}. The number of cases for isomorphism checking were  reduced by a large extent by obviating inter-class comparison through classification [Sec. \ref{2.3a}a]. 
\paragraph*{} Within each class, isomorphism among unlabeled decks had to be checked to see if they can uniquely identify a graph for all possible pairs of graphs, which constituted the most frequent step. Although such comparisons were reduced by a large number through classification of graphs, it still is the major contributor to the execution time in this module. To avoid this, the graphs in the decks are again classified on the basis of degree sequence. (Refer Sec.\ref{2.3b}b for details). The deck-isomorphism checking was done based on the structure represented in Sec. \ref{2.5}. 
\paragraph*{} At the lowest level of classification, graphs required for comparison were taken pairwise; hence dreadnaut interface to nauty \cite{b5} was used, for individual isomorphism-checking. The details of the structure used for this has been discussed in Sec. \ref{2.5}. The choice of Brendan McKay’s Nauty(No AUTomorphisms, Yes?)[B5] for use in our exhaustive verification is based on the experimental survey of Graph Isomorphism Algorithms in \cite{f1} which gives a comprehensive assessment of various GI algorithms which implement exact one-to-one matching using various techniques. Nauty was found to have the best performance time for small moderately dense graphs.

\subsection{\label{2.2} Complexity Issues }
\paragraph*{} For any class of graphs, containing say $n$ graphs, with $m$ vertices, the procedure requires comparing every pair ($^nC_2$ in this  case) for deck-level isomorphism, which in turn requires a worst case of $^{m+1}C_2$ isomorphism checks. In any iteration, two graphs are read from the file, and their decks are kept in main memory, each deck having $m$ graphs, stored as adjacency matrices.
\paragraph*{}
Time Complexity(Worst Case): \\  
           \hspace*{2 cm} $  ^nC_2 *\ ^{m+1}C_2 * O \{ GI$\footnote { The graph isomorphism problem is one of a very small number of problems belonging to NP neither known to be solvable in polynomial time nor NP-complete \cite{g1}, and a special complexity class $GI$ has been defined for such problems.}\}

\begin{figure*}
	  \centering 
           \includegraphics[width=18cm]{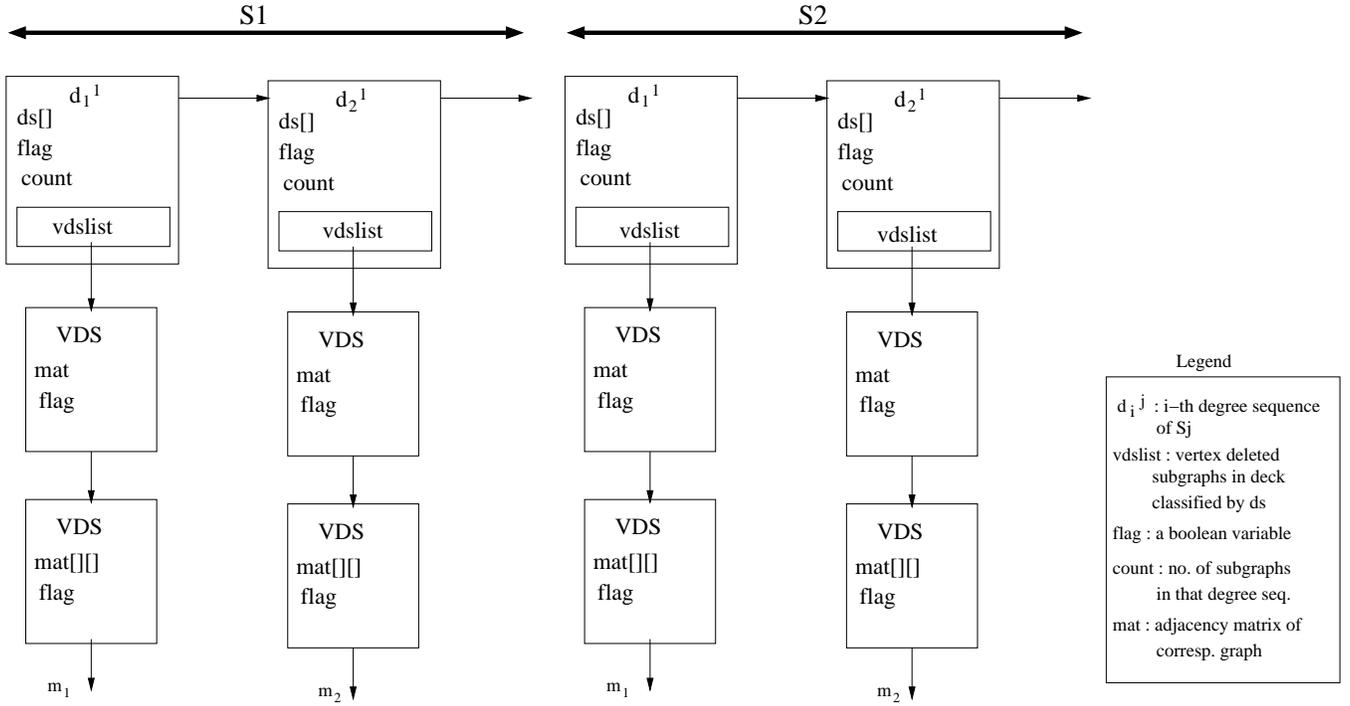}
	   \caption{The data structures used while comparing decks for isomorphism}
	   \label{deck}
  \end{figure*}

\subsection{\label{2.3} Pruning }
\paragraph*{} As discussed in the previous section, the complexity involved in checking RC for a certain class is very high as it involves  the comparison of the unlabeled decks of graph pairs. So, in order that lesser number of such comparisons are performed, the  graphs should be partitioned into mutually exclusive and exhaustive classes, so that inter-class graph pairs need no isomorphism checking. In our approach, pruning was employed  at two levels, where various parameters like degree sequence, characteristic polynomial, diameter were used to prune the graphs.  \\
\paragraph*{\label{2.3a} Level-1 } The listings of SC graphs were first classified according to degree sequences, so that any two graphs with different degree sequences don’t need to be compared. The divisions formed on basis of degree sequence were further classified into  groups  of graphs with the same characteristic polynomial. Two graphs in the same group   were compared only if their diameter was same. 

 Let the set of all classes thus formed be $S$, then the number of graph comparisons reduces as

\ \ \ \ \ \ \ \ \  $ ^{n}C_2 \rightarrow {\sum_{i \epsilon [\#S]}} ^{n_i}C_2 $

Fig. \ref{prune} depicts the improvements observed when all SC graphs over 16 graphs are pruned. The bars correspond to different classes with heights representing the no. of pairwise comparisons required within that class.
These pairs are passed on to the Graph Comparison module. The continuous curve intersecting the bars represents the actual number of pairs compared. \\ 

\paragraph*{\label{2.3b} Level-2} This level reduces number of isomorphism checks required in a graph comparison between some graphs $G_1$ and $G_2$. The graphs in corresponding decks $D1$ and $D2$ are classified according to degree sequences. 

The deck isomorphism is checked only if the number of graphs of any degree sequence in both the decks are equal. 

Let $S_i$ be the set of degree sequences of the graphs in $D_i$. Then the number of graph isomorphism checkings reduces as:

\ \ \ \ \ \ \ \  $ ^{m+1}C_2 \rightarrow {\sum_{i \epsilon [S_1 \cap S_2 ^{}]}} ^{m_i+1}C_2 $ \\

\subsection{\label{2.4} Analyzing the Numbers Involved }
\paragraph*{} As is clear from the plot shown [Fig. \ref{plot}],  the number of self complementary graphs rise steeply with the increase in the number of vertices. Thus, as one proceeds further, the storage as well as computation time becomes a dominant factor while analyzing the graphs. A detailed formula for the numbers of SC graphs can be found in the appendix. 
\paragraph*{} Additionally, as the number of vertices increase, the size of individual graph also increases. (equivalently the number of ascii characters required to encode an individual graph in graph6\cite{b3} format). 
\paragraph*{} Both these factors clearly indicate the sharp increase in the problem complexity on increasing the number of vertices. The approach employed to deal with both space and time factors have been have been discussed in further sections.

\subsection{\label{2.5} Storage and Implementation details }
\paragraph*{} The graphs were encoded into the graph6 format and stored as character strings in files. While comparing two graphs, the corresponding decks were stored in a structure as shown in Fig. \ref{deck}, to speed-up the deck-isomorphism checking. In case the two decks were concluded to be  different at any stage, the next graph pair was considered. If a pair of decks are found to be isomorphic, it can be concluded that the class under consideration is not uniquely reconstructible.
\paragraph*{} Each graph's deck has been stored as a list of its degree sequences comprising the vertex-deleted graphs. When the vertex-deleted subgraphs are being formed from the main graph, their degree sequences are calculated and they are appended to the existing structure. The number of subgraphs belonging to each degree sequence is stored as 'count', which is used as a measure of potential dissimilarity between the decks under inspection. 
\paragraph*{} The various terms used have been explained below: 
\begin{itemize}
\item $VDS$ (Vertex Deleted Subgraphs): to store a vertex deleted subgraph of the current graph i.e. one graph of the deck, along with its degree sequence.
\item $DS$ (Degree Sequence): to store together all the vertex deleted subgraphs of a graph that have the same degree sequence, and their count. 
\item $mat[][]$: represents the adjacency matrix of the graph. 
\item $ds[]$: is the degree sequence array for the graph. 
\end{itemize}

\section{ Results, Limiting Factors and  Future Work }
\paragraph*{} An exhaustive approach was followed in order to work towards disproving the conjecture for Self Complementary graphs. Since no counter example was found up to graphs on 17 points, the weak reconstruction was established for all SC graphs up to 17 vertices. 
\paragraph*{} The approach is limited by various factors. The number of SC graphs increases more than exponentially and generation and storage of these graphs becomes a problem. If we go by the way of exhaustive checking of RC for the whole class, the deck isomorphism checking for unlabeled graphs involves a large amount of computation(As discussed in Sec.\ref{2.2} , the isomorphism problem is neither Polynomial time, nor NP-Complete). Thus, progress using exhaustive approach is limited by the computational power. 
\paragraph*{} A possible approach could be to prune graphs on the basis of various properties to form classes, accounting for the rapid increase in numbers(Sec.\ref{2.4}). These classes are not necessarily mutually disjoint but jointly exhaustive, such that inter-class graph comparisons are not necessary, thereby reducing the amount of computation required, in terms of both space and time. Cluster computing or parallel programming can be used, but that can  take the endeavor only one step further. A search for a counterexample can end only when one has actually been found. 
\paragraph*{} Having validated the weak reconstruction of Self-Complementary graphs through exhaustive verification up to 17 vertices, proving  the Reconstruction Conjecture for SC Graphs requires the establishment of their Recognizability.

\newpage
\section* {Appendix} 
\paragraph*{} For any natural number $n$, there are no SC graphs on $4n+2$ and $4n+3$ vertices, \cite{r1}. Listing of no. of SC graphs up till 31 vertices is given by Sloane\cite{s3}.
 Table at the bottom gives the listing up till 101 vertices. It has been computed using the following formula given by \cite{r1}. \\
\paragraph*{} Let $\sigma_{k}$ be the number of SC graphs on k vertices,
$d(q,r)$ be the highest common factor of $q$ and $r$,
$(j)$ denotes summation for \\
$j_{1}+2j_{2}+3j_{3}+...+nj_{n} = n $ and $k_{s}=j_{4s}$,\\
then \\
\hspace*{1cm} $\sigma_{4N}$ = $\sum_{(k)}$ $\frac{2^{R}}{\prod_{s=1}^{N} s^{k_{s}}.k_{s}!}$ \\
where \\
 $R = 2\sum_{s=1}^{N}k_{s}(s k_{s}-1)+ 4\sum_{1\leq\alpha<\beta\leq N} k_{\alpha}k_{\beta} d(\alpha,\beta) $\\ \\
and \\
\hspace*{1cm} $\sigma_{4N+1}$ = $\sum_{(k)}$ $\frac{2^{R'}}{\prod_{s=1}^{N} s^{k_{s}}.k_{s}!}$ \\
where \\ 
$R' = \sum_{s=1}^{N}k_{s}(2s k_{s}-1)+ 4\sum_{1\leq\alpha<\beta\leq N} k_{\alpha}k_{\beta} d(\alpha,\beta) $\\

\newpage
\fontsize{10}{12} \selectfont
\begin{center} 
\begin{tabular}{ |p{1.1cm} | p{5.3cm} | p{1.1cm}|}
  \hline
  No. of vertices& No. of SC Graphs & No. of digits \\ \hline
1 & 1 & 1 \\ \hline
4 & 1 & 1 \\ \hline
5 & 2 & 1 \\ \hline
8 & 10 & 2 \\ \hline
9 & 36 & 2 \\ \hline
12 & 720 & 3 \\ \hline
13 & 5600 & 4 \\ \hline
16 & 703760 & 6 \\ \hline
17 & 11220000 & 8 \\ \hline
20 & 9168331776 & 10 \\ \hline
21 & 293293716992 & 12 \\ \hline
24 & 1601371799340544 & 16 \\ \hline
25 & 102484848265030656 & 18 \\ \hline
28 & 3837878966366932639744 & 22 \\ \hline
29 & 491247277315343649710080 & 24 \\ \hline
32 & 128777257564337108286016980992  & 30 \\ \hline
33 & 329669710587199326711682228592 64 & 32 \\ \hline
36 & 614548774973084626181885323304 10573824 & 38 \\ \hline
37 & 314648967511484697617766124367 41418123264 & 41 \\ \hline
40 & 422314689395950135433730499958 070655419345928192 & 48 \\ \hline
41 & 432450241375084625203842385525 712986695638650716160 & 51 \\ \hline
44 & 422127191316454227775483562647 79042838046660873019415068672 & 59 \\ \hline
45 & 864516487729608208102279735223 446130127050895825107423833620 48 & 62 \\ \hline
48 & 618845006889283567091123287966 324958874629599640950150338415 26613475328 & 71 \\ \hline 
49 & 205347891481994861280390367607 518623249435827483168144028565 4319521071104000 & 76 \\ \hline 
52 & 133991833678650709422844021098 422189855201413567485644125926 0891574069074668524929024 & 85 \\ \hline 
53 & 109766110223172826219628209432 329898030465321728087071079183 82132057567460666123980111872 & 89 \\ \hline 
56 & 431032551878974759633719601128 733652032040321118483003619093 693713491000324152998932957823 483510784 & 99 \\ \hline 
57 & 706203732998483706995227847596 308048394508719529822381377043 045445958522624420198995684123 8095158312960 & 103 \\ \hline 
60 & 207061124269499640717374782307 415600562906919880543389280838 917966133897871179404177048982 1148286755146482801704960 & 115 \\  
& & \\ \hline 
  
\end{tabular}

\begin{tabular}{ |p{1.1cm} | p{5.3cm} | p{1.1cm}|}
  \hline
61 & 678497892006294445422241289323 780963772966250550903395594719 662197061934532684847875764204 06501130798055917964390039552 & 119 \\ \hline 
64 & 149203264336484720095016247814 949538021562046303278771936611 682711188585667279315272321413 811185196875595690865500457245 516599656448 & 132 \\ \hline 
65 & 977818513155586058065330879779 815922582228413241077407349841 104091685122324386083797116636 681567831410799875295740469800 3213006781022208 & 136 \\ \hline 
68 & 161900848951598125922102268348 274406927651321450449665581675 617732454860248879488944014897 508474614218151466763396272067 235453567049272027118181548032  & 150 \\ \hline 
69 & 212206680737838692479598643071 561654840131716546507729167639 433940291992975403417181752868 466998761256350031595356886288 626360603761086478026590926772 63360 & 155 \\ \hline 
72 & 265470535637905592920486371298 931321812763903180076940405327 459440380849360413735101049838 607502256244469147877803424887 373170034661319568604549004911 1661980510574346240 & 169 \\ \hline 
73 & 695915080942631236783977078669 218752668281018309497989765767 218998018679614933199959215730 269451918584309173295516981430 082962221077114245985284385770 379081577130059748081664 & 174 \\ \hline 
76 & 659815347203130704718396676527 918606269829296456485444576342 226280280781022061748227086634 768047400194662842649127632935 670482530781830425344046435419 624192224493419861363519618091 086512128 & 189 \\ \hline 
77 & 345933268754420835150345813829 534665185861078612082569791333 115810000083686448692744624238 401839138403084821404189180884 505032830131842815959426804814 410030249788029524831615029411 122045482172416 & 195 \\ \hline 
80 & 249271190441194771178444224218 011427504063187636685521254468 806508744077622805751971945758 389779437805813670518550248581 082800838707387863208686111200 299708838138935425429444112542 015060564207753900206585806848 0 & 211 \\ \hline 

\end{tabular}

\begin{tabular}{ |p{1.1cm} | p{5.3cm} | p{1.1cm}|}
  \hline
  No. of vertices& No. of SC Graphs & No. of digits \\ \hline
& & \\
81 & 261379787788066248381893784432 358748599937185633737655731328 471298711235975024142007787467 498820630220524958537795800413 427774083069638061203812229145 778967002412151875405831456207 061407378384853940210846606727 4727424 & 217 \\ \hline 
84 & 143500180100207479114997189998 243862065540710190653850003333 819219400329763364736982700570 090472547380911009106852674830 812739206191280315369399095478 437471175257155465483947177581 332000048787390822621023311840 029743733810242498068480 & 234 \\ \hline 
85 & 300941689697510315240870035084 081885319109136917752219573706 986858476981024828585725084639 468252754069591552848951249304 303042586471904134865102515972 545201114567895301002357392115 989584048323232069154568737481 977621471324827215676259172352  & 240 \\ \hline 
88 & 126168052976356528017519464279 970380277246821645939532372117 786703617869491956079915156082 774256576345576135739799953970 896301446268644234425118898688 586793035588197774593492959263 752273662468394418895918966369 058683145825983584770262848456 903118303090704384 & 258 \\ \hline 
89 & 529187169270944090889981319556 369590445137244678795761778453 210660885257808496158528337839 545292996698799615169820977453 514666532275140114735433351477 361281588332949207084604669740 372448219564012348196733416859 665668093100517163412258909324 155360897097785281609728 & 264 \\ \hline 
92 & 169770091802294282623679980610 422948987480327508574748590219 251031687819084930652545336081 430303711088985854118477785758 308078696702827775637544817404 586921103270872330885802661077 601303715854098902152334825144 391186784527220692327515048706 803886992911981708632422630946 5545826631680 & 283 \\

& & \\ \hline 

\end{tabular}

\begin{tabular}{ |p{1.1cm} | p{5.3cm} | p{1.1cm}|}
  \hline
93 & 142413475025346023757126054636 489404110596945809534964721760 594089997064734029292487479224 694619622039351997502242950870 283643603609077989391316793869 702506201493004714417257399507 167816588476383580348334263512 643676670548312313904506066078 215275378061616438550112413902 00362471977318875136 & 290 \\ \hline 
96 & 350275323525357999981371792335 876318434320809029842859164255 150334412339862643411847434226 169395893161240502077884123858 229470450551749606322076202245 751917385494313364616822116714 785536672732266008431592758979 882511845456028102786108100130 928084285002917195332436745689 459703629357363973516979987822 046871552 & 309 \\ \hline 
97 & 587664476225481264301546988122 415642706567274358598705802054 533372768184097145478767168285 461794302703989616432536647317 677903892132394081205256893090 697563428694235217542235873400 144320903884185886519199233204 672316545493418732529695335942 097182349091099955758232428071 064964481182233998529310908521 5730542026162176 & 316 \\ \hline 
100 & 111006681028014327834610597428 047356644234351893275341023004 385295174188465093716126926615 428068369513879188836930958823 298533800885315689471621833687 008582437384982360781801876927 495343430023442155870301467400 904001588267394665516380953759 067688589315526078816735090650 884591327071497291715624135415 245761938051150481982141016616 3636224 & 337 \\ \hline 
101 & 372476613010019685835214850967 089506247560906619265669939557 238645582991681329809801908265 460289453555648405462061297321 886991960883229387325315792582 383510469111367735641248273012 069213497083312610889264458676 142645287501471643006531219109 082039861359541440703399104224 499941550106700669434764941849 254397291031046851000833746631 99397546819584 & 344 \\ \hline 

\end{tabular}

\end{center}

\end{document}